\documentclass[%
preprint,
showkeys,
aps,
nofootinbib %force the footnot presents in the same page.
 ]{revtex4-1}
\usepackage{graphicx}% Include figure files
\usepackage{dcolumn}% Align table columns on decimal point
\usepackage{float} % 重要：引入float包
\usepackage{lipsum} % 仅y用于生成示例文本
\usepackage{hyperref}% add hypertext capabilities
\hypersetup{
  colorlinks=true,
  linkcolor=blue,
  filecolor=magenta,
  urlcolor=cyan,
  citecolor=cyan,
}
\usepackage{amsmath,amsfonts,amssymb}
\usepackage{array}
\usepackage{braket} %please add this package we need it.
\usepackage{color}  % for color
\usepackage{xcolor}
\usepackage{subfiles}
\usepackage{comment}
\begin{document}

%\begin{comment}
  \title{\huge Wallis Products from the Four-Dimensional Singular Harmonic Oscillator\\
  \vspace{1cm} \color{teal}{\hrule height 3.0pt} }% \vspace{0.5cm}\includegraphics[width=5.3cm]{circle-4.jpg}}% Force line breaks with \\

%\begin{center}
%  \includegraphics[width=2.3cm]{circle-5.jpg}
%\end{center}
%\end{comment}

%\title{\huge Wallis Products from the Four-Dimensional Singular Harmonic Oscillator}

\author{Bin Ye}
\author{Ruitao Chen}
\author{Lei Yin}
\email[Corresponding Author: ]{lei@scnu.edu.cn}
\affiliation{School of Materials and New Energy, Xingzhi College, South China Normal University, Shanwei 516625, China.}
\affiliation{ State Key Laboratory of Nuclear Physics and Technology, Institute of Quantum Matter, South China Normal University, Guangzhou 510006, China}
%\altaffiliation[Also at ]{South China Normal University}


\begin{abstract}
    We present a variational derivation of the Wallis product and its reciprocal from the four-dimensional singular harmonic oscillator. The inverse-square interaction is absorbed into an effective angular parameter $\nu$, so that the lowest exact energy in a fixed sector is $E_{4d,\mathrm{exact}}=\hbar\omega(\nu+2)$. Motivated by the radial Kustaanheimo--Stiefel relation $r=\rho^2$ between the four-dimensional oscillator and the three-dimensional Coulomb problem, we use the quartic trial family $R_a(\rho)=N\rho^\nu e^{-a\rho^4}$. The minimized variational energy yields an accuracy ratio governed by adjacent Gamma functions. In the large-$\nu$ semiclassical limit, this ratio approaches unity. Restricting $\nu$ to the odd sequence $\nu=2n-1$ gives the standard Wallis product, whereas the even sequence $\nu=2n$ gives its reciprocal form. The Coulomb-dual interpretation further relates the two branches to integer and half-integer effective angular sectors in the dual Coulomb/MICZ description. The result shows that Wallis-type infinite products persist under an inverse-square deformation of the oscillator and arise from a common Gamma-function structure in radial variational dynamics.
\end{abstract}

%Use showkeys class option if keyword

%- make the contents disappear before submission.
\maketitle
\tableofcontents

\section{Introduction}

The Wallis product
\begin{equation}
\frac{\pi}{2}
=
\prod_{k=1}^{\infty}
\frac{(2k)^2}{(2k-1)(2k+1)}
\label{eq:Wallis_product}
\end{equation}
is a classical infinite product for $\pi$ with a long history in analysis \cite{Wallis1655,BerggrenBorweinBorwein1997}. Standard derivations are usually based on trigonometric integrals, Gamma-function identities, or asymptotic estimates. A surprising quantum-mechanical route to Eq.~\eqref{eq:Wallis_product} was proposed by Friedmann and Hagen, who showed that the Wallis product can be recovered from a variational estimate of the hydrogen-atom energy in the large-angular-momentum limit \cite{FriedmannHagen2015}. In their construction, the variational energy contains a ratio of Gamma functions with adjacent integer and half-integer arguments, and the correspondence-principle limit converts this Gamma-function ratio into the Wallis product.

Subsequent work clarified both the scope and the limitations of this observation. Chashchina and Silagadze showed that the appearance of the Wallis product is not tied uniquely to the Gaussian trial function used in the original hydrogenic calculation; rather, the relevant Wallis ratio is already encoded in the radial integrals and may also emerge from other admissible trial families \cite{ChashchinaSilagadze2017}. Cortese and Garc\'ia further demonstrated that the hydrogen atom is not the only quantum system in which such a structure appears. Using the Coulomb--oscillator duality, they related the Wallis formula to a four-dimensional harmonic oscillator with a quartic exponential trial function \cite{CorteseGarcia2018}. These results suggest that the essential mechanism is not a single special Hamiltonian, but a combination of radial variational calculus, Gamma-function moments, and a semiclassical large-angular-momentum limit.

The purpose of the present work is to examine this mechanism in a singular radial system. We consider the four-dimensional singular harmonic oscillator with potential
\begin{equation}
V(\rho)
=
\frac{1}{2}m\omega^2\rho^2+\frac{g}{\rho^2},
\qquad \rho>0 ,
\label{eq:intro_singular_potential}
\end{equation}
where the inverse-square term modifies the effective centrifugal barrier. Singular inverse-square interactions are a standard testing ground for the short-distance structure of quantum mechanics, because they probe the balance between angular momentum, confinement, and possible collapse \cite{Case1950,deAlfaroFubiniFurlan1976}. In the present radial problem, the singular coupling can be absorbed into an effective angular parameter $\nu$, defined by
\begin{equation}
\nu(\nu+2)
=
L(L+2)+\frac{2mg}{\hbar^2}.
\label{eq:intro_effective_nu}
\end{equation}
The lowest energy in a fixed effective sector is then $\hbar\omega(\nu+2)$. Thus the singular oscillator extends the ordinary four-dimensional oscillator by replacing the hyperspherical quantum number $L$ with a coupling-dependent effective parameter $\nu$.

Our variational construction is guided by the Kustaanheimo--Stiefel relation between the four-dimensional oscillator and the three-dimensional Coulomb problem \cite{KustaanheimoStiefel1965}. Under the radial map $r=\rho^2$, a Gaussian factor of the form $\exp(-ar^2)$ in the Coulomb radial problem is transformed into the quartic factor $\exp(-a\rho^4)$ in the oscillator variable. We therefore use the trial family
\begin{equation}
R_a(\rho)
=
N\rho^\nu e^{-a\rho^4},
\qquad a>0 ,
\label{eq:intro_trial_function}
\end{equation}
where the short-distance factor $\rho^\nu$ incorporates the effective inverse-square behavior. The resulting variational energy can be evaluated analytically and gives the dimensionless ratio
\begin{equation}
R_{4d}(\nu)
=
\frac{E_{\mathrm{var}}(\nu)}
{E_{4d,\mathrm{exact}}(\nu)}
=
\frac{\sqrt{2(\nu+3)}}{\nu+2}
\frac{\Gamma\!\left(\frac{\nu+3}{2}\right)}
{\Gamma\!\left(\frac{\nu+2}{2}\right)}.
\label{eq:intro_ratio}
\end{equation}
The large-$\nu$ limit gives $R_{4d}(\nu)\to1$, reflecting the asymptotic exactness of the variational state in the correspondence-principle regime.

Equation~\eqref{eq:intro_ratio} is the central object of the paper. When $\nu$ is restricted to the odd sequence $\nu=2n-1$, the ratio contains the Gamma channel $\Gamma(n+1)/\Gamma(n+1/2)$ and yields the standard Wallis product. When $\nu$ is restricted to the even sequence $\nu=2n$, it contains the complementary channel $\Gamma(n+3/2)/\Gamma(n+1)$ and yields the reciprocal Wallis product,
\begin{equation}
\prod_{k=1}^{\infty}
\frac{(2k-1)(2k+1)}{(2k)^2}
=
\frac{2}{\pi}.
\label{eq:intro_reciprocal_Wallis}
\end{equation}
The singular oscillator therefore packages the standard and reciprocal Wallis products into two complementary integer subsequences of the same effective-angular-momentum ratio.

We also discuss the Coulomb-dual interpretation of this construction. In the dual three-dimensional description, the effective parameter $\lambda=\nu/2$ plays the role of an angular parameter in a Coulomb-type radial problem. This is naturally compatible with the integer and half-integer angular sectors that occur in MICZ--Kepler systems with Dirac-quantized monopole charge \cite{Dirac1931,McIntoshCisneros1970,Zwanziger1968,Lavrenov2019}. In this sense, the singular oscillator provides a compact setting in which inverse-square dynamics, Coulomb--oscillator duality, and Wallis-type Gamma ratios meet within a single variational framework.

The paper is organized as follows. In Sec.~2 we formulate the four-dimensional singular oscillator and evaluate the quartic variational energy. In Sec.~3 we derive the standard and reciprocal Wallis products from the odd and even effective-angular sectors. In Sec.~4 we give the Coulomb-dual interpretation and summarize the result.

\section{Singular oscillator and quartic variational family}

We consider the four-dimensional singular harmonic oscillator in a fixed hyperspherical angular-momentum sector. Writing the wave function as $\Psi(\rho,\Omega)=R(\rho)Y_{L\alpha}(\Omega)$, where $Y_{L\alpha}$ is a hyperspherical harmonic on $S^3$ satisfying
\begin{equation}
-\Delta_{S^3}Y_{L\alpha}(\Omega)
=
L(L+2)Y_{L\alpha}(\Omega),
\qquad L=0,1,2,\ldots ,
\label{eq:S3_harmonic}
\end{equation}
the radial Hamiltonian is
\begin{equation}
\hat H_L
=
-\frac{\hbar^2}{2m}
\left(
\frac{\mathrm{d}^2}{\mathrm{d}\rho^2}
+\frac{3}{\rho}\frac{\mathrm{d}}{\mathrm{d}\rho}
-\frac{L(L+2)}{\rho^2}
\right)
+\frac{1}{2}m\omega^2\rho^2+\frac{g}{\rho^2}.
\label{eq:radial_H_L}
\end{equation}
The inverse-square interaction may be absorbed into an effective angular parameter $\nu$ through
\begin{equation}
\nu(\nu+2)
=
L(L+2)+\frac{2mg}{\hbar^2}.
\label{eq:nu_definition}
\end{equation}
Equivalently,
\begin{equation}
\nu
=
-1+\sqrt{(L+1)^2+\frac{2mg}{\hbar^2}},
\label{eq:nu_branch}
\end{equation}
where we take the regular branch. Reality of $\nu$ requires
\begin{equation}
g \geq
-\frac{\hbar^2}{2m}(L+1)^2 \,.
\label{eq:g_reality_bound}
\end{equation}
The stronger restriction $\nu\geq0$, equivalent to $g\geq-\hbar^2L(L+2)/(2m)$, selects the regular effective sectors used in the following variational construction. With Eq.~\eqref{eq:nu_definition}, the Hamiltonian becomes:
\begin{equation}
\hat H_{\nu}
=
-\frac{\hbar^2}{2m}
\left(
\frac{\mathrm{d}^2}{\mathrm{d}\rho^2}
+\frac{3}{\rho}\frac{\mathrm{d}}{\mathrm{d}\rho}
-\frac{\nu(\nu+2)}{\rho^2}
\right)
+\frac{1}{2}m\omega^2\rho^2 .
\label{eq:effective_H_nu}
\end{equation}
Thus the singular oscillator is reduced to an ordinary four-dimensional radial oscillator with a shifted angular parameter. The exact energy spectrum in this effective sector is
\begin{equation}
E_{n_r,\nu}
=
\hbar\omega(2n_r+\nu+2),
\qquad n_r=0,1,2,\ldots ,
\label{eq:exact_spectrum}
\end{equation}
and the lowest energy at fixed $\nu$ is therefore
\begin{equation}
E_{4d,\mathrm{exact}}(\nu)
=
\hbar\omega(\nu+2).
\label{eq:exact_ground_energy}
\end{equation}

We now introduce the quartic variational family
\begin{equation}
R_a(\rho)
=
N\rho^\nu e^{-a\rho^4},
\qquad a>0 .
\label{eq:quartic_trial}
\end{equation}
The power $\rho^\nu$ gives the regular short-distance behavior associated with the effective inverse-square barrier, while the quartic exponential is motivated by the radial Kustaanheimo--Stiefel relation $r=\rho^2$: a Gaussian Coulomb trial factor $e^{-ar^2}$ is pulled back to $e^{-a\rho^4}$ in the oscillator coordinate.

The normalization condition is
\begin{equation}
\int_0^\infty \rho^3 |R_a(\rho)|^2\,\mathrm{d}\rho
=
1 .
\label{eq:normalization_condition}
\end{equation}
Using Eq.~\eqref{eq:quartic_trial}, one obtains
\begin{equation}
|N|^2
=
\frac{4(2a)^{(\nu+2)/2}}
{\Gamma\!\left(\frac{\nu+2}{2}\right)} .
\label{eq:normalization_constant}
\end{equation}
The moments of $\rho$ are then
\begin{equation}
\langle \rho^q\rangle
=
(2a)^{-q/4}
\frac{
\Gamma\!\left(\frac{\nu}{2}+1+\frac{q}{4}\right)
}{
\Gamma\!\left(\frac{\nu}{2}+1\right)
}.
\label{eq:rho_moments}
\end{equation}

Acting with the effective Hamiltonian \eqref{eq:effective_H_nu} on the trial function gives
\begin{equation}
\hat H_{\nu}R_a
=
\left[
\frac{\hbar^2}{2m}
\left(
8a(\nu+3)\rho^2-16a^2\rho^6
\right)
+\frac{1}{2}m\omega^2\rho^2
\right]R_a .
\label{eq:H_on_trial}
\end{equation}
Therefore the variational energy is
\begin{equation}
\langle H\rangle(a)
=
\frac{\hbar^2}{2m}
\left[
8a(\nu+3)\langle \rho^2\rangle
-16a^2\langle \rho^6\rangle
\right]
+\frac{1}{2}m\omega^2\langle \rho^2\rangle .
\label{eq:energy_a}
\end{equation}
It is convenient to introduce
\begin{equation}
y
=
\sqrt{2a}.
\label{eq:y_definition}
\end{equation}
Substituting the moments \eqref{eq:rho_moments} into Eq.~\eqref{eq:energy_a}, the energy functional becomes
\begin{equation}
\langle H\rangle(y)
=
\frac{
\Gamma\!\left(\frac{\nu+3}{2}\right)
}{
\Gamma\!\left(\frac{\nu+2}{2}\right)
}
\left[
\frac{\hbar^2}{m}(\nu+3)y
+\frac{m\omega^2}{2y}
\right].
\label{eq:energy_y}
\end{equation}
The minimum occurs at
\begin{equation}
y_\ast
=
\frac{m\omega}{\hbar\sqrt{2(\nu+3)}} ,
\label{eq:y_star}
\end{equation}
and the minimized variational energy is
\begin{equation}
E_{\mathrm{var}}(\nu)
=
2\hbar\omega
\sqrt{\frac{\nu+3}{2}}
\frac{
\Gamma\!\left(\frac{\nu+3}{2}\right)
}{
\Gamma\!\left(\frac{\nu+2}{2}\right)
}.
\label{eq:E_var}
\end{equation}
Comparing this result with the exact ground-state energy \eqref{eq:exact_ground_energy}, we obtain the accuracy ratio
\begin{equation}
R_{4d}(\nu)
=
\frac{E_{\mathrm{var}}(\nu)}
{E_{4d,\mathrm{exact}}(\nu)}
=
\frac{\sqrt{2(\nu+3)}}{\nu+2}
\frac{
\Gamma\!\left(\frac{\nu+3}{2}\right)
}{
\Gamma\!\left(\frac{\nu+2}{2}\right)
}.
\label{eq:R4d}
\end{equation}
Since $E_{\mathrm{var}}(\nu)$ is a variational upper bound for the lowest energy in the fixed effective sector, $R_{4d}(\nu)\geq1$. Moreover, using the asymptotic relation \cite{TricomiErdelyi1951,DLMFGamma}  $\Gamma(z+1/2)/\Gamma(z)\sim z^{1/2}$ as $z\to\infty$, Eq.~\eqref{eq:R4d} gives
\begin{equation}
\lim_{\nu\to\infty}R_{4d}(\nu)
=
1 .
\label{eq:R4d_limit}
\end{equation}
This large-$\nu$ limit is the correspondence-principle input that will convert the Gamma-function ratio in Eq.~\eqref{eq:R4d} into the Wallis products in the next section.

\section{Odd and even Wallis channels}

The ratio \eqref{eq:R4d} contains two complementary Gamma-function channels, depending on whether the effective angular parameter $\nu$ is restricted to odd or even integer values. For fixed hyperspherical quantum number $L$, imposing an integer value of $\nu$ is equivalently a discrete choice of the inverse-square coupling,
\begin{equation}
g(\nu)=\frac{\hbar^2}{2m}\left[\nu(\nu+2)-L(L+2)\right].
\end{equation}
The odd and even subsequences considered below should therefore be understood as two arithmetic families of effective singular couplings.
 We first consider the odd sequence
\begin{equation}
\nu=2n-1,
\qquad n=1,2,\ldots .
\label{eq:odd_nu}
\end{equation}
Substitution into Eq.~\eqref{eq:R4d} gives
\begin{equation}
R_{4d}^2(2n-1)
=
\frac{4n+4}{(2n+1)^2}
\left[
\frac{\Gamma(n+1)}
{\Gamma\!\left(n+\frac{1}{2}\right)}
\right]^2 .
\label{eq:R4d_odd_square}
\end{equation}
Using
\begin{equation}
\Gamma(n+1)=n!,
\qquad
\Gamma\!\left(n+\frac{1}{2}\right)
=
\frac{(2n-1)!!}{2^n}\sqrt{\pi},
\qquad
(2n)!!=2^n n!,
\label{eq:gamma_double_factorial_odd}
\end{equation}
we obtain
\begin{equation}
\left[
\frac{\Gamma(n+1)}
{\Gamma\!\left(n+\frac{1}{2}\right)}
\right]^2
=
\frac{[(2n)!!]^2}
{\pi[(2n-1)!!]^2}.
\label{eq:gamma_ratio_odd}
\end{equation}
Therefore
\begin{equation}
R_{4d}^2(2n-1)
=
\frac{4n+4}{(2n+1)^2}
\frac{[(2n)!!]^2}
{\pi[(2n-1)!!]^2}.
\label{eq:R4d_odd_double_factorial}
\end{equation}
The finite Wallis product is
\begin{equation}
W_n
=
\prod_{k=1}^{n}
\frac{(2k)^2}{(2k-1)(2k+1)}
=
\frac{[(2n)!!]^2}
{(2n+1)[(2n-1)!!]^2}.
\label{eq:finite_Wallis_product}
\end{equation}
Hence Eq.~\eqref{eq:R4d_odd_double_factorial} becomes
\begin{equation}
R_{4d}^2(2n-1)
=
\frac{4n+4}{\pi(2n+1)}W_n .
\label{eq:R4d_odd_Wn}
\end{equation}
Taking the limit $n\to\infty$ and using Eq.~\eqref{eq:R4d_limit}, we find
\begin{equation}
1
=
\frac{2}{\pi}
\lim_{n\to\infty}W_n .
\label{eq:odd_limit_relation}
\end{equation}
Thus the odd effective-angular sequence yields the standard Wallis product
\begin{equation}
\prod_{k=1}^{\infty}
\frac{(2k)^2}{(2k-1)(2k+1)}
=
\frac{\pi}{2}.
\label{eq:standard_Wallis_from_odd}
\end{equation}

We next consider the even sequence
\begin{equation}
\nu=2n,
\qquad n=1,2,\ldots .
\label{eq:even_nu}
\end{equation}
For this branch, Eq.~\eqref{eq:R4d} gives
\begin{equation}
R_{4d}^2(2n)
=
\frac{4n+6}{(2n+2)^2}
\left[
\frac{\Gamma\!\left(n+\frac{3}{2}\right)}
{\Gamma(n+1)}
\right]^2 .
\label{eq:R4d_even_square}
\end{equation}
Using
\begin{equation}
\Gamma\!\left(n+\frac{3}{2}\right)
=
\frac{(2n+1)!!}{2^{n+1}}\sqrt{\pi},
\qquad
\Gamma(n+1)=n!,
\qquad
(2n)!!=2^n n!,
\label{eq:gamma_double_factorial_even}
\end{equation}
we have
\begin{equation}
\left[
\frac{\Gamma\!\left(n+\frac{3}{2}\right)}
{\Gamma(n+1)}
\right]^2
=
\frac{\pi[(2n+1)!!]^2}
{4[(2n)!!]^2}.
\label{eq:gamma_ratio_even}
\end{equation}
It is useful to introduce the reciprocal finite Wallis product
\begin{equation}
\widetilde W_n
=
\prod_{k=1}^{n}
\frac{(2k-1)(2k+1)}{(2k)^2}
=
\frac{(2n+1)[(2n-1)!!]^2}
{[(2n)!!]^2}.
\label{eq:finite_reciprocal_Wallis_product}
\end{equation}
Since $(2n+1)!!=(2n+1)(2n-1)!!$, Eq.~\eqref{eq:gamma_ratio_even} can be written as
\begin{equation}
\left[
\frac{\Gamma\!\left(n+\frac{3}{2}\right)}
{\Gamma(n+1)}
\right]^2
=
\frac{\pi(2n+1)}{4}\widetilde W_n .
\label{eq:gamma_ratio_even_Wtilde}
\end{equation}
Substituting this result into Eq.~\eqref{eq:R4d_even_square}, we obtain
\begin{equation}
R_{4d}^2(2n)
=
\frac{\pi(4n+6)(2n+1)}
{4(2n+2)^2}\widetilde W_n .
\label{eq:R4d_even_Wtilde}
\end{equation}
Taking $n\to\infty$ and again using Eq.~\eqref{eq:R4d_limit}, the rational prefactor tends to $\pi/2$, and therefore
\begin{equation}
1
=
\frac{\pi}{2}
\lim_{n\to\infty}\widetilde W_n .
\label{eq:even_limit_relation}
\end{equation}
Thus the even effective-angular sequence gives the reciprocal Wallis product
\begin{equation}
\prod_{k=1}^{\infty}
\frac{(2k-1)(2k+1)}{(2k)^2}
=
\frac{2}{\pi}.
\label{eq:reciprocal_Wallis_from_even}
\end{equation}

Equations~\eqref{eq:standard_Wallis_from_odd} and \eqref{eq:reciprocal_Wallis_from_even} show that the standard and reciprocal Wallis products arise from two complementary integer subsequences of the same variational ratio \eqref{eq:R4d}. The odd branch contains the Gamma ratio $\Gamma(n+1)/\Gamma(n+1/2)$, whereas the even branch contains its complementary shifted ratio $\Gamma(n+3/2)/\Gamma(n+1)$. In this sense, the singular oscillator packages the two Wallis channels into a single effective-angular-momentum structure.

\section{Coulomb-dual interpretation and conclusion}

The derivation in the previous sections was carried out entirely within the four-dimensional singular oscillator. It is nevertheless useful to interpret the result through the Coulomb--oscillator duality, because this explains why the quartic trial family \eqref{eq:quartic_trial} is a natural one rather than a purely algebraic choice. The radial Kustaanheimo--Stiefel map relates the three-dimensional Coulomb coordinate $r$ and the four-dimensional oscillator coordinate $\rho$ by
\begin{equation}
r=\rho^2 .
\label{eq:KS_radial_map}
\end{equation}
Under this map, a Gaussian Coulomb trial factor $e^{-ar^2}$ is transformed into the quartic oscillator factor $e^{-a\rho^4}$. At the radial level, the absorbed effective-barrier description identifies the corresponding Coulomb angular parameter as
\begin{equation}
\lambda=\frac{\nu}{2}.
\label{eq:lambda_nu_relation}
\end{equation}
Thus the trial state \eqref{eq:quartic_trial} is the oscillator-side image of the Coulomb-type radial family
\begin{equation}
R_a^{(C)}(r)
=
\mathcal N r^\lambda e^{-ar^2},
\qquad a>0 .
\label{eq:Coulomb_trial}
\end{equation}

Following the same Coulomb variational calculation used in Ref.~\cite{FriedmannHagen2015}, but with an effective angular parameter $\lambda$, one obtains the dimensionless ratio
\begin{equation}
R_{3d}(\lambda)
=
\frac{(\lambda+1)^2}{\lambda+\frac{3}{2}}
\left[
\frac{\Gamma(\lambda+1)}
{\Gamma\!\left(\lambda+\frac{3}{2}\right)}
\right]^2 .
\label{eq:R3d_lambda}
\end{equation}
Substituting Eq.~\eqref{eq:lambda_nu_relation} into Eq.~\eqref{eq:R3d_lambda}, one obtains 
\begin{equation}
R_{3d}(\nu)
=
\frac{(\nu+2)^2}{2(\nu+3)}
\left[
\frac{\Gamma\!\left(\frac{\nu+2}{2}\right)}
{\Gamma\!\left(\frac{\nu+3}{2}\right)}
\right]^2 .
\label{eq:R3d_nu}
\end{equation}
Comparison with Eq.~\eqref{eq:R4d} gives the simple inverse relation
\begin{equation}
R_{3d}(\nu)
=
\frac{1}{R_{4d}^2(\nu)} .
\label{eq:R3d_R4d_relation}
\end{equation}
Here both the variational and exact Coulomb energies are negative, so that the ratio is positive.

This relation is a compact expression of the dual nature of the two variational problems. On the oscillator side, the quartic trial function gives an upper bound to the positive ground-state energy, so that $R_{4d}(\nu)\geq1$. On the Coulomb side, the corresponding variational estimate is less negative than the exact binding energy, and therefore the positive ratio $R_{3d}(\nu)$ is less than or equal to unity. Both ratios approach one in the same large-$\nu$ limit.

The same interpretation also clarifies the role of the odd and even branches. In the ordinary Coulomb problem, the angular quantum number is integral. In the MICZ--Kepler system, where a charged particle moves in the field of a Coulomb center and a magnetic monopole, the Dirac quantization condition gives
\begin{equation}
\mu\in\frac{1}{2}\mathbb Z .
\label{eq:Dirac_quantization}
\end{equation}
The corresponding angular sectors are naturally of the form
\begin{equation}
\lambda=|\mu|+q,
\qquad q=0,1,2,\ldots .
\label{eq:MICZ_angular_sector}
\end{equation}
Hence $\lambda$ may be either integer or half-integer, and Eq.~\eqref{eq:lambda_nu_relation} implies that $\nu=2\lambda$ belongs to an integer sequence. The even sequence $\nu=2n$ corresponds to integer $\lambda$, while the odd sequence $\nu=2n-1$ corresponds to half-integer $\lambda$. Therefore, the two Wallis channels found in Sec.~3 are naturally compatible with the integer and half-integer angular sectors of the Coulomb-dual MICZ setting. This observation should be understood as a physical interpretation of the branch structure, not as an additional assumption required for the variational derivation itself.

We have shown that the four-dimensional singular harmonic oscillator provides a compact variational route to both the Wallis product and its reciprocal. The inverse-square interaction shifts the effective angular parameter from $L$ to $\nu$, while the Kustaanheimo--Stiefel radial map motivates the quartic trial family $R_a(\rho)=N\rho^\nu e^{-a\rho^4}$. The resulting variational accuracy ratio is governed by adjacent Gamma functions,
\begin{equation}
R_{4d}(\nu)
=
\frac{\sqrt{2(\nu+3)}}{\nu+2}
\frac{\Gamma\!\left(\frac{\nu+3}{2}\right)}
{\Gamma\!\left(\frac{\nu+2}{2}\right)} ,
\label{eq:R4d_summary}
\end{equation}
and the large-$\nu$ semiclassical limit $R_{4d}(\nu)\to1$ converts this ratio into infinite products. The odd sequence $\nu=2n-1$ produces the standard Wallis product, whereas the even sequence $\nu=2n$ produces the reciprocal product.

The significance of this result is threefold. First, it shows that the Wallis structure is stable under an inverse-square deformation of the oscillator, provided that the singular term is absorbed into an effective angular parameter. Second, it unifies the standard and reciprocal Wallis products as two complementary sectors of the same Gamma-function ratio. Third, it connects the variational origin of the Wallis product with the Coulomb--oscillator duality and with the integer/half-integer angular sectors familiar from monopole quantum mechanics. In this sense, the present construction adds the singular oscillator to the family of quantum-mechanical systems in which classical infinite-product identities emerge from radial variational dynamics and their semiclassical limits.

%%%%%%%%%%%%%%%%%%%%%%%%%%%
%\bibliographystyle{JHEP}
%\bibliography{SigularWallisReferences}

\providecommand{\href}[2]{#2}\begingroup\raggedright\endgroup

\end{document}